\documentclass[pre,reprint,aps]{revtex4-1}
\usepackage{amsmath}
\usepackage{amssymb}
\usepackage{graphicx}% Include figure files
\usepackage{dcolumn}% Align table columns on decimal point
\usepackage{bm}% bold math
\usepackage{hyperref}% add hypertext capabilities

\newcommand{\PD}{{\mbox{PD}}}
\newcommand{\Per}{{\mathbf{Per}}}
\newcommand{\Dev}{{\mathbf{Dev}}}

\newcommand{\seg}{\,\mbox{s}}
\newcommand{\cm}{\,\mbox{cm}}

\begin{document}

\title{Dynamics of silo deformation under granular discharge}
\author{Claudia Colonnello}
\affiliation{Departamento de F\'isica, Universidad Sim\'on Bol\'ivar, AP 89000, Caracas 1080-A, Venezuela}
\author{Miroslav Kram\'ar}
\affiliation{INRIA Saclay,
1 Rue Honoré d'Estienne d'Orves, 91120 Palaiseau, France}

\date{\today}

\begin{abstract}

We use Topological Data Analysis to study the post buckling behavior of laboratory scale cylindrical silos under gravity driven granular discharges.
Thin walled silos buckle during the discharge if the initial height of the granular column is large enough. The deformation of the silo is reversible as long as the filling height does not exceed a critical value, $L_c$. Beyond this threshold the deformation becomes permanent and the silo often collapses.
We study the dynamics of reversible and irreversible deformation processes, varying the initial filling height around $L_c$.
We find that  all reversible processes exhibit striking similarities and they alternate between regimes of slow and fast dynamics.
The patterns that occur at the beginning of irreversible deformation processes are topologically very similar to those that arise during reversible processes. However, the dynamics of reversible and irreversible processes is significantly different. In particular, the evolution of irreversible processes is much faster. This allows  us to make an early prediction of the collapse of the silo based solely on observations of the deformation patterns. 

\end{abstract}


\maketitle

\section{Introduction}
\label{sec::intro}

Thin walled cylindrical silos are used extensively as storage facilities for granular materials in industrial applications.
Despite great efforts to develop appropriate design and construction protocols, their structural failure remains a widespread problem~\cite{rotter08, dogangun09, dutta13}, leading to important economic losses and risks for human personnel.

One of the main factors hindering the efficient prevention of silo failure is the difficulty in accounting for the interaction between the grains and the silo~\cite{rotter08}, due to the complex stress patterns that arise within confined granular media (e.g~\cite{liu95, majmudar05, vanel99b}).
The model proposed by Janssen \cite{janssen} provides a simple description of the stress on a silo filled with a granular material and is widely used as a reference in engineering applications~\cite{nedderman, rotter08}. It accurately describes the total stress on the walls and bottom of a silo if local contact shear forces exerted by the grains are in a state of maximal mobilization~\cite{vanel99, ovarlez03, bertho03}. However, the state of mobilization is strongly affected by the details of the granular packing and the filling procedure~\cite{vanel99, back11}. During a granular discharge of the silo the maximal mobilization condition is attained spontaneously if it is not initially satisfied~\cite{perge12}. Once the maximal mobilization condition is met, the total stress during the discharge is well described by the model~\cite{perge12, cambau13}. 

Another limitation of Janssen's model is that it fails to predict local stress fluctuations on the walls of the silo, as it is based on a description of the granular material as a continuum. Local stress measurements and simulations show that fluctuations can be very large, particularly at the onset of a granular discharge~\cite{wang15, wang13}. Moreover, their distribution may not be symmetric in the angular direction~\cite{zhong01, ostendorf03}. 
Such stress fluctuations are likely to induce geometric defects  and  load eccentricities on the silo wall, that can affect the stability of the structure. In fact, the critical buckling load of a thin empty shell exhibits a strong imperfection sensitivity~\cite{horton65, simitses86}, which represents an enduring challenge for the accurate description of the shell's stability~\cite{jansseune16, skukis17, virot17}.

\begin{figure}
\includegraphics{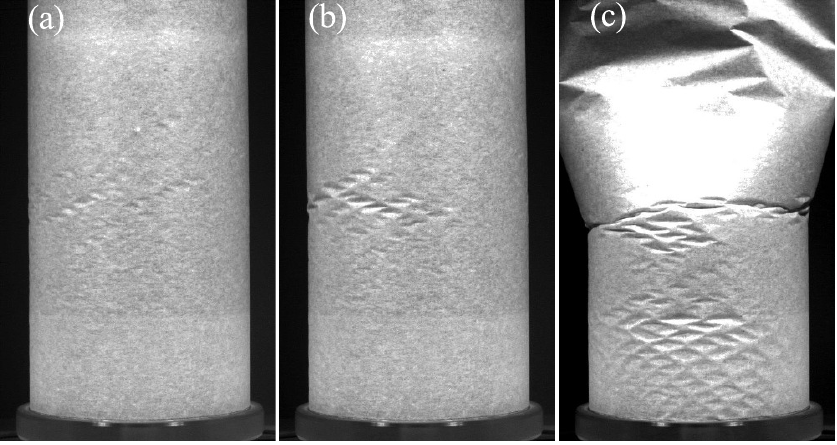}
\caption{Snapshots of deformation patterns observed in our experiments.
Using the qualitative criteria of Sec.~\ref{sec::results::qualitative}, the corresponding deformation processes are classified as
(a) far-from-collapse, (b) close-to-collapse, and (c) collapse processes.
(a) and (b) show the shell when the deformation stops growing, right before it starts returning to its original undeformed state.
(c) shows a shell  after catastrophic failure. The upper section of the shell has collapsed against a protective glass cylinder surrounding the silo.}
\label{fig::classes-images}
\end{figure}

The most common failure mode for silos with thin and flexible vertical walls is buckling due to the effective axial compression resulting from frictional forces exerted by the grains \cite{rotter08}. 
In particular, if the initial filling height of the silo is large enough, its vertical walls buckle during the gravity driven discharge of the granular material~\cite{gutierrez15, colonnello14}. 
During the discharge, diamond shaped indentations develop at the lower end of the cylindrical shell and form intricate deformation patterns (see Fig.~\ref{fig::classes-images}). The number and size of the indentations are strongly affected by local stress fluctuations due to granular collisions and they tend to increase with the initial filling height of the silo.  
If the filling height exceeds a critical value, $L_c$, then the indentations grow large enough to produce a permanent deformation of the shell. This is usually followed by a catastrophic collapse of the structure. Otherwise, all indentations disappear completely by the end of the granular discharge. The scaling behavior of $L_c$ has been studied systematically in~\cite{gutierrez15}, where it is successfully described by a model that considers the stability of an empty shell subject to an effective axial stress derived from Janssen's model.  
However, the evolution of the deformation patterns is far from being understood.

Local features of the stress distribution on the silo wall, such as fluctuations and friction mobilization, are expected to play a fundamental role in the deformation process~\cite{gutierrez15}. However, direct local stress measurements on thin walled silos can be challenging. A grid of strain sensors has been used on large scale experiments~\cite{ostendorf03, gallego15, wang15}, but such methods are difficult to implement at laboratory scale, and could affect significantly the response of very thin shells. These difficulties have been circumvented by considering the deformation patterns that develop on the shell as it interacts with the granular material~\cite{zhao13, cambau13}. For small shell deformations, this allows for the reconstruction of the stress field on the shell. 

In this paper we present a quantitative study of the dynamics of post buckling deformation patterns induced by granular discharges on thin walled cylindrical silos.
We use images obtained from laboratory scale experiments using paper silos, and employ Topological Data Analysis (TDA) to study the patterns that develop on the silo while varying the initial filling height around the critical value, $L_c$.  
We use persistent homology\cite{edelsbrunner-harer,carlsson} to represent a deformation process as a trajectory in the space of persistence diagrams. This space is well suited to capture important topological information about the pattern and quotient out the overwhelming geometrical variations~\cite{PhysD2016}. In this way, we compare different deformation processes and gain novel insights into their evolution, that contribute to our understanding of the conditions that lead to the failure of the structure. 

We find that deformation patterns at early stages of irreversible processes leading to the collapse of the silo are very similar to the patterns exhibited by reversible processes. However, already at these early stages their dynamics  is significantly different. In particular, irreversible processes evolve faster. This allows us to effectively predict the collapse at an early stage of the discharge.
The existence of such early differences between reversible and irreversible processes indicates that the collapse is triggered by partially mobilized grain-to-wall friction, as proposed in~\cite{colonnello14,gutierrez15}.
We also observe evidence of slow-fast dynamics in the evolution of the deformation and argue that it is caused by the intrinsic nonlinear behavior of the thin shell. 
 
The paper is organized as follows. Section~\ref{sec::methods} gives a detailed account of our methods. In Sec.~\ref{sec::results} we present our results.
First, in Sec.~\ref{sec::results::qualitative} we introduce a set of qualitative criteria to classify a deformation process after the discharge. We define three classes according to the deformation sustained by the silo. 
In Secs.~\ref{sec::results::dynamics}~and~\ref{sec::Reversibility} we analyze the dynamics of deformation processes in each of these classes, and discuss  the evidence of fast-slow dynamics as well as the reversibility of the trajectories.  We compare the trajectories for all deformation processes in Sec.~\ref{sec::CrossCompare} and analyze quantitative differences and similarities between trajectories in each class. 
In Sec.~\ref{sec::classification} we discuss the early classification of the processes.
Finally, in Sec.~\ref{sec::FT} we discuss the origin of the fast-slow dynamics. 
We conclude with some final remarks in Sec.~\ref{sec::conclusions}.

\section{Methods}\label{sec::methods}
\subsection{Experimental setup} 
\label{sec::methods::exp-methods}

\begin{figure}
\includegraphics{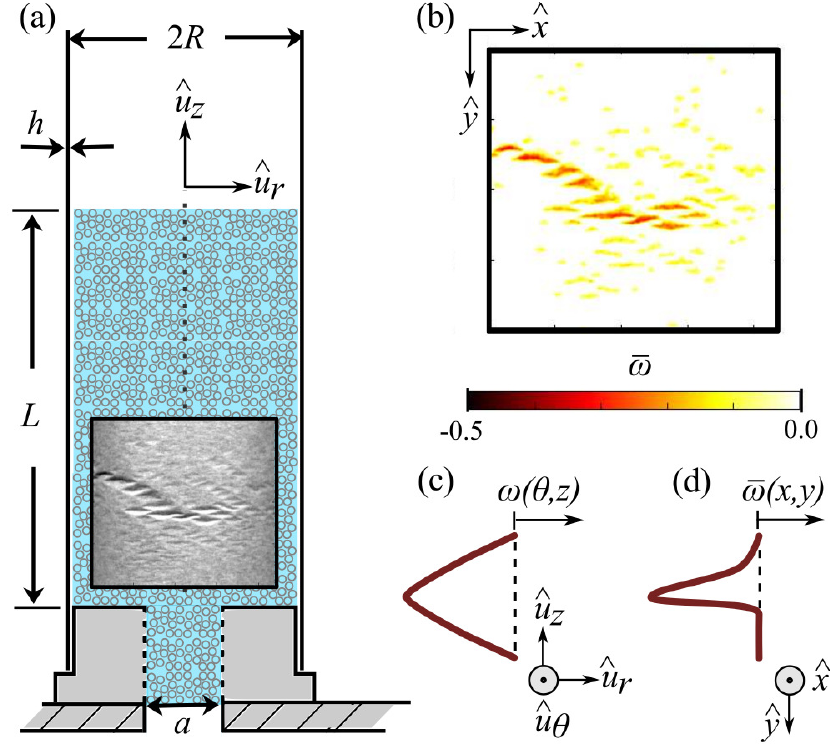}
\caption{(Color online) (a) Schematic cross-section of the experimental setup indicating its dimensions. $L$ indicates the height of the granular column. 
A typical image used in our analysis is shown, marking the area of the shell it captures.
Due to oblique illumination from above, the upper (lower) parts of diamond shaped indentations produce dark (bright) features in the image. 
(b) Field $\bar\omega$ obtained from processing the image of the deformation pattern shown in (a). 
(c) Schematic representation of 
$\omega(\theta, z)$, for fixed $\theta$, of the cross-section at the center of a single diamond shaped indentation in the Yoshimura pattern.
Referential directions corresponding to cylindrical coordinates $(r, \theta, z)$ on the shell are indicated.
Here $\theta=\theta(x,y)$ and $z=z(x,y)$, with $(x,y)$ Cartesian coordinates on the image, as indicated in (d). 
(d) Schematic representation of $\bar\omega(x,y)$, corresponding to the cross-section of the indentation shown in (c).}
\label{fig::SetUp}
\end{figure}

Figure \ref{fig::SetUp}(a) shows a sketch of the experimental setup. 
Silos used in our experiments are built following the protocol described in \cite{gutierrez15}. 
A cylindrical paper shell, with thickness $h=(24\pm5)\,\mu\mbox{m}$ and radius $R=(4.0\,\pm\,0.1)\cm$, is fitted tightly on a solid acrylic base fixed to a level surface. The upper boundary of the silo remains free.
The solid base has a centered orifice of diameter $a=(0.5\,\pm\,0.1)\cm$. We start by closing the  orifice  with a foam plug. 
Then the silo is filled at constant flow rate with glass beads of diameter $d=(1.50\,\pm\,0.05)\,\mbox{mm}$, through a fixed funnel centered above the silo. After filling the silo, the plug is removed and the gravity driven granular discharge starts.

The critical filling height for our silos (measured as described in \cite{gutierrez15}) is $L_c=(11.5\,\pm\,0.6)\cm$.
To estimate the effect of fluctuating deformations induced by the granular flow, we perform one granular discharge with an initial filling height $L_o^f=(7.0\,\pm\,0.3)\cm$.  This filling height is sufficiently below $L_c$ to guarantee that the silo wall only exhibits deformations due to collisions with the moving particles. 
The remaining granular discharges are performed with initial filling heights $L_o>0.8L_c$, so that persistent deformation patterns  develop as well  on the silo wall.
We study five different silos and a total of 27 granular discharges. We progressively increase $L_o$ in successive granular discharges of a single silo until an evident permanent deformation develops on its wall. 

In order to record the configuration of the deformation patterns, the silo is obliquely illuminated by a strong light source from above. In this way, the typical diamond shaped indentations are clearly visible as pairs of approximately symmetric dark and bright features, see Fig.~\ref{fig::SetUp}(a). 
A digital camera placed in front of the silo is used to capture a sequence of images during the granular discharge. The sampling starts simultaneously with the discharge and continues for $8.3\,s$.
The duration of the sampling is set by limited hardware resources and does not encompass the whole granular discharge.
However, this has only a minor impact on our results, as our main interest lies in the emergence and early development of the deformation. Moreover, if a permanent deformation occurs it is always before $8.3\,s$ and we truncate the sequence shortly after we detect it by visual inspection. 

Grayscale digital images with size $440\times400\,\mbox{pix}$ are collected at a fixed rate of $30$ frames per second, i.e. we take an image at every time $t_i = i/30\,s$ for $0\leq i \leq250$. Images capture  a fixed area of the shell spanning roughly $100^o$ of its circumference, as depicted in Fig.~\ref{fig::SetUp}(a).

\subsection{Image processing} 

To study the topological features of the deformation patterns and their evolution during a granular discharge, we construct a time series of dimensionless scalar fields $\{\bar\omega_i\}$ for each discharge.  
A piecewise constant scalar field $\bar\omega_i$ is obtained from the digital image of the shell taken at time $t_i$ (see Sec.~\ref{sec::methods::exp-methods}).
In the rest of this section we will describe how to construct $\bar\omega$ at a fixed time from an image of the shell. Then we will show that $\bar\omega$ retains relevant information about the topology of the deformation pattern exhibited by the shell at this time.

First, the intensity field of the digital image is normalized by a median filtered (size=50pix) reference image taken right before the onset of the discharge. This  compensates for inhomogeneities in the illumination of the shell partly caused by its curvature. The reference image may exhibit some incipient deformations produced during the filling of the silo but their size is marginal and their effect on the median filtered intensity field is negligible. The normalized intensity field defined on the pixels of the original image is denoted by $I$.

If a point $(x,y)$  in the image belongs to a dark feature, corresponding to the upper part of a diamond shaped indentation, then $I(x,y) < 1$. Conversely, for  $I(x,y) > 1$ the point corresponds to the bottom part of the indentation.
We retain only information about the upper part of the indentations to simplify the procedure and define
\[
\bar\omega(x,y) = \mbox{min}\{ I(x,y), 1 \} - 1.
\]
Thanks to the characteristic symmetry of the indentations, considering only their upper part is sufficient, as we will show below.
Finally, we apply an erosion filter on $\bar\omega(x,y)$ to remove small noisy features produced by the texture of the paper.
An example of an original image of the pattern and the corresponding field $\bar\omega$ resulting from this procedure are shown in Figs.~\ref{fig::SetUp}(a) and (b), respectively.

The shape of the deformed shell is completely described by the radial displacement from its original cylindrical state, $\omega$. 
The fields $\omega$ and $\bar\omega$ are far from being identical. 
However, we will show that their similarities within the set $\bar\Omega_0 = \{ (x,y) \colon \bar\omega(x,y) < 0\}$ make 
$\bar\omega$ a good descriptor of the topology of the deformation pattern.

The definition of $\bar\omega$ and the illumination of the silo ensure that for a paper sheet folded to recreate the Yoshimura pattern \cite{yoshimura55}, there exists a monotone function $f\colon \mathbb R \to  \mathbb R$ such that $|f(\omega(x,y)) - \bar\omega(x,y)| \leq 0.07 f(\omega(x,y))$ for $(x,y) \in \bar\Omega_0$. 
Figures \ref{fig::SetUp}(c)-(d) show the fields $\omega$ and $\bar\omega$ on a cross-section of an indentation from the Yoshimura pattern.
The folded shell is made with the same paper and has the same radius as the silos used in our experiments. 
The resulting regular indentations are very similar in size and shape to those that develop during the discharge.  
Therefore, we expect that in our experiments the differences between $\bar\omega$ and $f(\omega)$ have the same bound as in the case of the Yoshimura pattern.

Finally, we show that $\bar\omega$ is a good descriptor of the topology of the deformation pattern. In particular, consider the set of indentations  given by  $\{ (x,y) \colon \omega(x,y) < 0 \}$. The geometry of this set differs considerably from the geometry of $\bar\Omega_0$ but their topology is closely related. For every diamond shaped indentation, its upper part is contained in  $\bar\Omega_0$, so for every indentation there is a connected component in $\bar\Omega_0$. Moreover, if two indentations merge they always do so along an edge, as the vertices exhibit an increased stiffness thanks to their increased curvature~\cite{pini16}. 
Thus, the connected components in $\bar\Omega_0$ corresponding to these indentations merge into a single one as well.

These observations support the existence of a close relation between the topology of the sub-level set
\[
\bar\Omega_\gamma = \{ (x,y) \colon \bar\omega(x,y) < \gamma \}
\]
and the set $\{ (x,y) \colon \omega(x,y) < \gamma \}$. This justifies  our choice to study the topological features of the pattern represented by $\bar\omega$.

\subsection{Dynamics in the Space of Persistence Diagrams}
\label{sec::Methods::PD}

Persistent homology \cite{edelsbrunner-harer,carlsson} is used to encode topological features of the pattern given by a scalar field $\bar\omega$. In particular, two persistence diagrams $\PD_0(\bar\omega)$ and $\PD_1(\bar\omega)$ are assigned to $\bar\omega$  . These  diagrams describe changes in topology of the sub-level sets  $\bar\Omega_\gamma$, for all $\gamma \in \mathbb R$.
Every point $(b,d) \in \PD_0(\bar\omega)$ ($\PD_1(\bar\omega)$)  encodes the appearance  of a connected component (loop) of $\bar\Omega_\gamma$ at $\gamma = b$ and its disappearance at $\gamma = d$. The lifespan of $(b,d)$, defined by $d -b$, is often interpreted as a measure of prominence of the topological feature  represented by $(b,d)$.
For an in-depth presentation describing the connection between the persistence diagrams and the underlying pattern we refer the reader to~\cite{PhysD2016}. 

More classical measures, such as number  and maximal depth of indentations, are readily available from the persistence diagrams. Due to the characteristic diamond shape of buckling indentations, even the total area they cover can be estimated. Conversely, some of our results can be obtained by analyzing the evolution of the area covered by indentations during the discharge.

There are well defined notions of distance between pairs of persistence diagrams~\cite{PhysD2016}. We will only use the so called Wasserstein distance, $d_{W^2}$.
This distance measures the overall difference between the persistence diagrams, encoding the difference between the patterns they represent.
As such, it measures changes in both the number and depth of the indentations. Larger changes in depth contribute more to the distance between the corresponding diagrams. 

The set of all persistence diagrams equipped with the $d_{W^2}$ distance is a complete metric space denoted by $\Per$. This, combined with the fact that the persistence diagrams are good descriptors of patterns, makes $\Per$ a good choice of an observation space. 
For  a given discharge, a trajectory $X \colon [0,\infty) \to \Per$ describes the evolution of the observed deformation pattern. By using the methods presented above we produce a sample of this trajectory 
\[
X(t_i) = (\PD_0(\bar\omega_i) , \PD_1(\bar\omega_i) ), 
\]
at times $t_i = i/30\,s$, for $0\leq i \leq250$.

\section{Results and discussion}
\label{sec::results}

\subsection{Qualitative classification}
\label{sec::results::qualitative}

Observation of the pattern through the granular discharge allows for a qualitative classification of the deformation processes into three classes:
(\emph{i)}  \emph{Far-from-collapse} (FFC), (\emph{ii)} \emph{Close-to-collapse} (CTC), and (\emph{iii)} \emph{Collapse} (C) processes.  A collapse process is one that results in a permanent deformation of the shell, detectable by visual inspection after the granular discharge has concluded. These processes often lead to a catastrophic collapse of the silo. They  occur if $L_o > L_c$ and always produce indentations  that grow until they exceed the elastic limit of the shell.  
 
Conversely, if $L_o<L_c$ a \emph{noncollapse} deformation process occurs, i.e., all visible deformations completely disappear by the end of the discharge. We say a noncollapse process is far-from-collapse if all indentations remain comparable in size to the fluctuating deformations induced by the granular flow. Since these deformations are produced by collisions of the grains against the silo wall they exhibit small values of the radial displacement of the shell ($\omega \lesssim 10^{-2}R$). Nevertheless, large clusters of localizations may develop during a FFC process. 
If a noncollapse process exhibits indentations that attain a larger radial displacement before receding, we say it is a close-to-collapse process. 
Figure \ref{fig::classes-images} shows characteristic examples of the deformation pattern for each class. In the rest of this paper we will demonstrate quantitative differences and similarities between the three classes.

\subsection{Dynamics of individual deformation processes}
\label{sec::results::dynamics}

\begin{figure}
\includegraphics{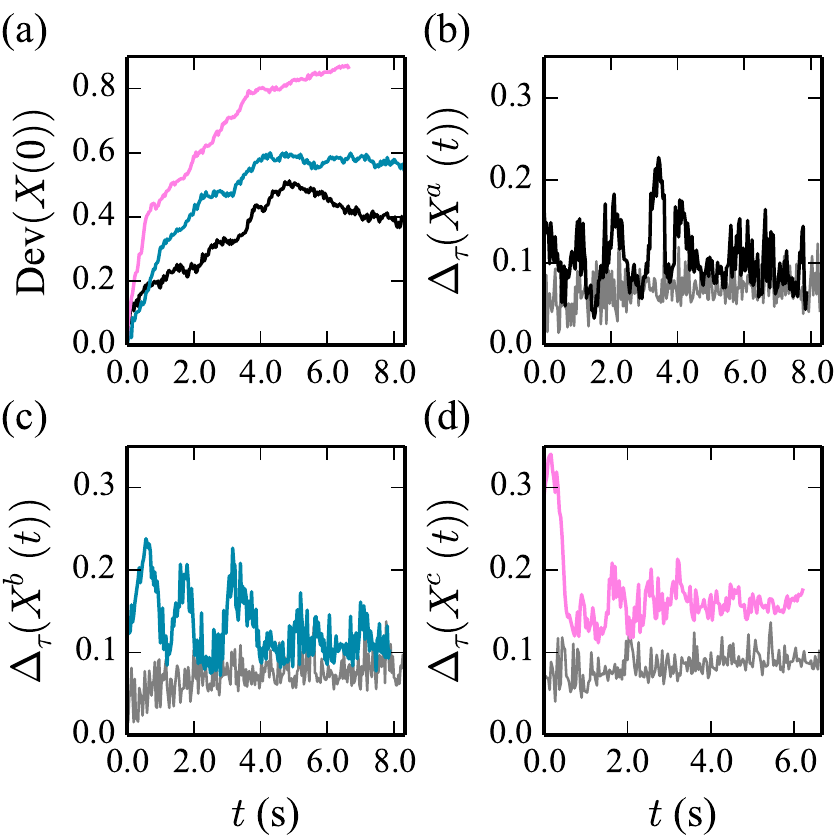}
\caption{(Color online) (a) Deviation from the initial condition along typical trajectories representing FFC, CTC and C deformation processes $X^a$ (black line), $X^b$ (blue line), and $X^c$ (pink line), respectively. 
(b)-(d) Magnitude of the displacement for $X^a$, $X^b$ and $X^c$ at two different time scales, $\tau=0.03\seg$ (gray line) and $\tau=0.43\seg$ (lines colored as in (a)).}
\label{fig::first_row} 
\end{figure}

In this section we consider the trajectories (in $\Per$)  of typical deformation processes in each of the qualitative classes introduced in Sec.~\ref{sec::results::qualitative}. For the sake of clarity we only consider  trajectories corresponding  to three successive discharges of the same silo, with filling heights $L_o^a=(11.2\,\pm\,0.3)\cm$ (FFC), $L_o^b=(11.5\,\pm\,0.3)\cm$ (CTC), and $L_o^c=(11.6\,\pm\,0.3)\cm$ (C).  In Sec.~\ref{sec::CrossCompare}, we show that trajectories of processes in the same class exhibit strong quantitative similarities.  To simplify the notation, we  denote the trajectories of the representative FFC, CTC, and C processes  by  $X^a, X^b$, and $X^c$, respectively.

We start by studying the deviation of a trajectory $X$ from its initial condition $X(0)$, which represents 
the marginal deformation pattern sustained by the shell at the onset of the discharge. 
Figure~\ref{fig::first_row}(a) shows  that for all three trajectories the distance from the initial condition, $\Dev(X(t)) := d_{W^2}(X(0),X(t))$,  initially increases. This reflects the growth and proliferation of indentations at the beginning of the discharge.

As expected, the deviation from the initial condition is smallest for the  FFC trajectory $X^a$ and largest for  the  C trajectory $X^c$. Notice that both $\Dev(X^a(t))$ and $\Dev(X^b(t))$ stop increasing around $t \approx 4\seg$. 
In noncollapse processes all indentations stop growing before the elastic limit of the shell is exceeded and they disappear completely by the end of the granular discharge. So, the trajectories $X^a$ and $X^b$ eventually return close to their initial condition. Indeed,  $\Dev(X^a(t))$ starts decreasing at $t = 4.86\seg$ as the trajectory starts returning towards $X^a(0)$. The restricted size of our sample does not allow us to observe the decrease in  $\Dev(X^b(t))$ but its value starts fluctuating around its maximum as indentations stop growing. Conversely, $\Dev(X^c(t))$ exhibits  a sustained increase, reflecting the ongoing proliferation and growth of indentations that leads to the collapse of the silo.

Another important property of the trajectory is how fast it evolves. In order to asses this, we define the magnitude of the displacement on the trajectory $X$ at time $t$ and time scale $\tau$ by
\[
\Delta_{\tau}(X(t))= d_{W^2}(X(t),X(t+\tau)). 
\]
Figures~\ref{fig::first_row}(b)-(d) show that for a short time scale, $\tau=0.03\seg$, the values of $\Delta_{\tau}(X^a(t))$, $ \Delta_{\tau}(X^b(t))$, and $\Delta_{\tau}(X^c(t))$ are very similar. 
Moreover, they are comparable with fluctuating deformations of the shell induced by collisions during the granular flow. To estimate the magnitude of these fluctuations, we consider the trajectory $X^f \subset \Per$  of a discharge  with a low initial filling height, $L_o^f$, for which only deformations induced by collisions are present (see Sec.~\ref{sec::methods::exp-methods}). 
The maximum value of $\Delta_{0.03}(X^f(t))$ is $0.1$ and we interpret it as the magnitude of deviations caused by flow induced fluctuations along the trajectory of any deformation process.
Once larger persistent indentations appear on the shell they are likely to favor collisions and thus increase the level of fluctuations. However, interactions between persistent and fluctuating deformations are expected to be complex~\cite{gutierrez15} and their analysis is out  of the scope of the present paper.

For time scales longer than $\tau \approx 0.40\seg$, $\Delta_{\tau}(X)$ exhibits a more interesting behavior. Remarkably, 
the noncollapse trajectories $X^a$ and $X^b$ tend to evolve slower than $X^c$, as corroborated by  $\Delta_{0.43}(X(t))$ shown in  Figs.~\ref{fig::first_row}(b)-(d). After  $t \approx 4\seg$,  the values of $\Delta_{0.43}(X^{a,b}(t))$  and $\Delta_{0.03}(X^{a,b}(t))$ are comparable, indicating that  the evolution of the systems is dominated by fluctuations \cite{kondic2017evolution}. This does not come as a surprise considering the behavior of $\Dev(X^{a,b}(t))$ for $t > 4\seg$.

The curve $\Delta_{0.43}(X^c(t))$, depicted in Fig.~\ref{fig::first_row}(d), indicates that $X^c$ evolves fastest soon after the onset of the discharge and slows down around $t \approx 1\seg$. This slowdown causes a drop in the slope of $\Dev(X^c(t))$.
The slope of $\Dev(X^c(t))$ changes once again around  $t \approx 4\seg$ but with no corresponding change in $\Delta_{0.43}(X^c(t))$. This implies that, instead of slowing down, at this time the trajectory $X^c$ changes the direction in which it evolves. 
Visual inspection of the pattern shows that this coincides with a significant change in the dynamics of the deformation. 
Namely, the  growth of individual indentations almost stops and the subsequent changes of the pattern are predominantly due to the merging of neighboring indentations. Merging constitutes an important growth mechanism for indentations, which individually do not overcome a characteristic size determined by the properties of the shell \cite{colonnello14}. In this way,  very large indentations that cause the collapse of a silo can develop.

Figures~\ref{fig::first_row}(c)-(d) show that for $t < 4\seg$ the curves $\Delta_{0.43}(X^{a,b}(t))$ exhibit a sequence of pronounced maxima and minima. 
Thus, at the time scale $\tau = 0.43\seg$, both $X^a$ and $X^b$ alternate between regimes of slow and fast dynamics.
In the rest of this section we will study this phenomenon in detail.
To show that it is not an artifact caused by a particular choice of timescale, we introduce a distance matrix that provides a comprehensive description of the dynamics at all available time scales. 
We will also show how to detect the transitions between slow and fast regimes using the distance matrix. The results obtained will prove fundamental for our analysis of the origin of these transitions described in Sec~.\ref{sec::FT}.

\begin{figure*}
\includegraphics{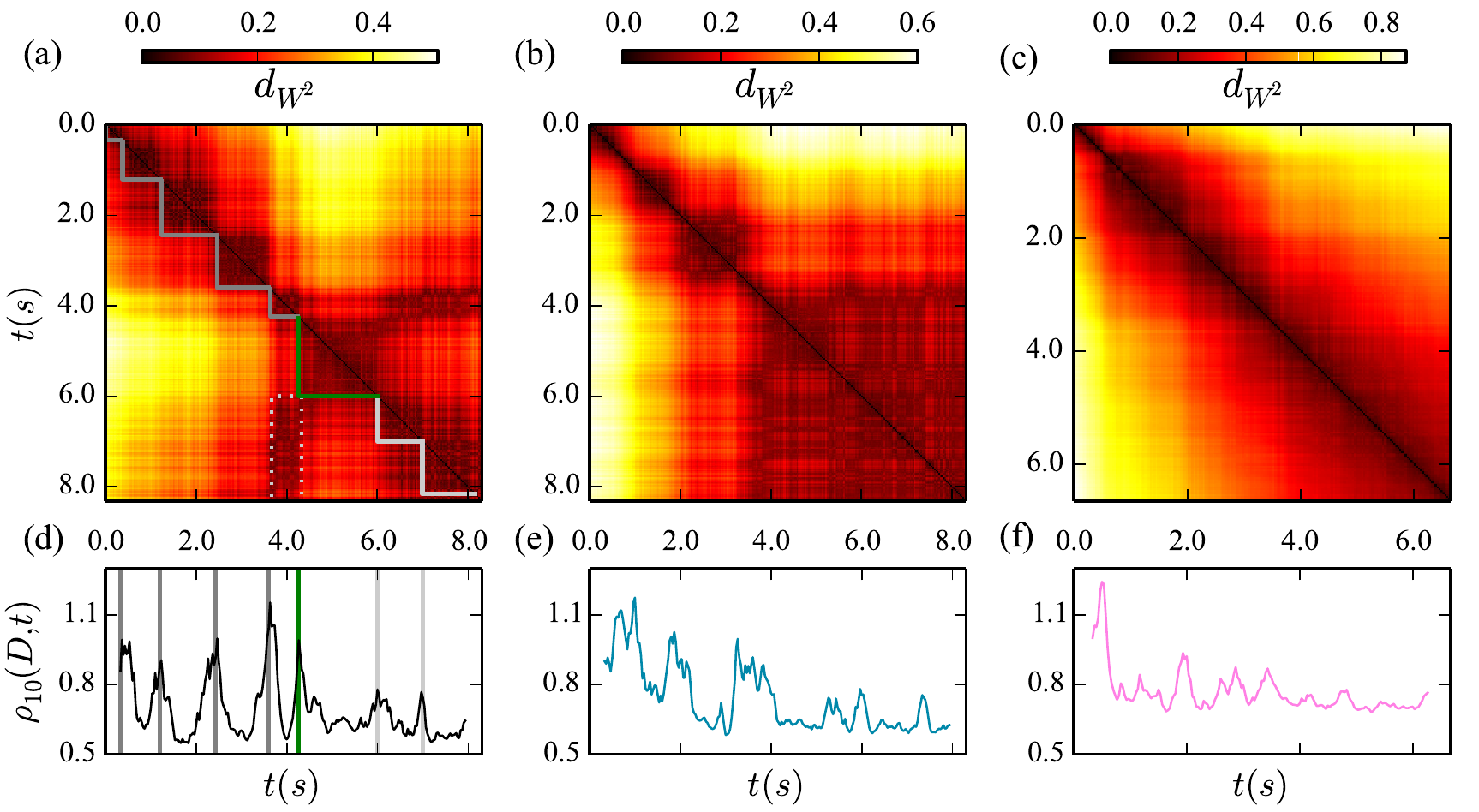}
\caption{(Color online) Distance matrices  (a) $D^a$, (b) $D^b$ and (c) $D^c$ for the trajectories $X^a$ (FFC), $X^b$ (CTC) and $X^c$ (C), respectively. 
Ratio (d) $\rho_{10}(D^a, t)$, (e) $\rho_{10}(D^b, t)$, and (f) $\rho_{10}(D^c, t)$.
In (d), vertical lines mark the maxima of $\rho_{10}(D^a, t)$, corresponding to fast transitions in the evolution of $X^a$. The green line marks the last forward transition of $X^a$, i.e., its onset of return time, $t_r$. Dark and light gray lines correspond to the remaining forward and backward transitions of $X^a$. In (a) boundaries of  diagonal blocks exhibiting slow dynamics are identified as transition times marked in (d), with corresponding line coloring. The dotted gray rectangular line marks an off diagonal block within $D^a$ with small values,  which signals the onset of the return of $X^a$ towards its initial condition.}
\label{fig::matrices} 
\end{figure*}

Let $X(t)$ in  $\Per$ be a trajectory sampled at times $t_i$, for $0 \leq i \leq n-1$, then the distance matrix of $X$ is an $n\times n$ matrix defined by 
\[
D(i,j) = d_{W^2}(X(t_i),X(t_j)).
\]
If there is no possibility of confusion, we denote $D(i,j)$ by $D(t_i,t_j)$ to indicate the time corresponding to the samples $i,j$. In this notation $D(t_i,t_j)$ is the $d_{W^2}$ distance between points $X(t_i)$ and $X(t_j)$. It quantifies the magnitude of the displacement on the trajectory between times $t_i$ and $t_j$. So, $\Delta_{\tau}(X(t_i)) = D(t_i,t_i+\tau)$ and  $\Dev(X(t_i)) = D(0, t_i)$.

Figures~\ref{fig::matrices}(a)-(c) show distance matrices $D^a, D^b$ and $D^c$ for the representative trajectories $X^a, X^b$ and $X^c$, respectively. These matrices exhibit different patterns, revealing interesting features of the dynamics of the underlying trajectories. We stress that similar patterns are observed in all matrices corresponding to the same class.

Now we analyze the dynamics of the trajectory $X^a$ by using its distance matrix. 
In Fig.~\ref{fig::matrices}(a)  one distinguishes a sequence of distinct dark blocks along the diagonal of $D^a$. 
Inside of each block the evolution of  $X^a$ is slow. To be more precise, if the times $t_i$ and $t_j$ fall in a single block, then the distance $D^a(t_i,t_j)$ is of the order of the estimated value of fluctuations. Note that the separation between distinct dark blocks is rather sharp. This means that during a short time interval around the boundary between two successive blocks the dynamics suddenly becomes fast.

To detect transition points along a trajectory $X$, represented by a distance matrix $D$, we consider the following ratio inspired by Fisher's discriminant~\cite{fisher1936use},
 \[
\rho_w(D, t_i)\,=\,\frac{\sum\limits_{n,m=0}^{w}D(t_{i-n},t_{i+m})}{\sum\limits_{n,m=0}^{w}(D(t_{i-n},t_{i-m}) +D(t_{i+n},t_{i+m}))},
\]
defined on a sliding window with length $2w$. If the time interval between two consecutive fast transitions of $X$ is longer than $w\Delta t$, and the duration of a single fast transition is short with respect to $w\Delta t$, then $\rho_w(D,t)$ attains prominent local maxima around the times at which these transitions take place.  
Thus, these time scales determine the choice of $w$ for an optimal detection of the fast transitions. One must also consider that if $w$ is too small fluctuations might compromise the results. Accordingly, we choose $w \in [10,25]$, which corresponds to time scales between $0.3\seg$ and $0.8\seg$. We find that the results are almost identical within this interval, thus we report the results only for $w = 10$. 

For the FFC trajectory $X^a$, Fig.~\ref{fig::matrices}(d) shows sharp peaks of $\rho_{10}(D^a,t)$ with well defined maxima indicating abrupt transitions. The positions of the maxima are used to obtain the boundaries of the blocks, drawn in Fig.~\ref{fig::matrices}(a), separating  time  intervals during which the evolution of the orbit is slow. 
Also CTC trajectories exhibit such prominent transitions, as illustrated in Figs.~\ref{fig::matrices}(b)~and~\ref{fig::matrices}(e) for $X^b$. However, CTC trajectories exhibit a faster evolution than FFC trajectories at the onset of the discharge  (see Figs.~\ref{fig::first_row}(a)-(c)). As a result, the peaks exhibited by $\rho_{10}(D^b,t)$ are wider and more rugged, as shorter time scales are required to differentiate some transitions. 

The curve $\rho_{10}(D^c, t)$ for the C trajectory $X^c$,  shown in Fig.~\ref{fig::matrices}(f), is different from the curves for noncollapse processes.
Its values tend to be larger and its peaks are less prominent, with a single exception soon after the onset of the discharge. The initial peak of $\rho_{10}(D^c,t)$ is caused by the rapid slowdown of the system soon after the onset of the discharge. Therefore, at time scales larger than $0.3\seg$ there is no evidence of an alternation between slow and fast dynamics in C processes. This is in accord with the behavior of $\Delta_{0.43}(X^c(t))$ shown in Fig.~\ref{fig::first_row}(d). Nevertheless, slow-fast dynamics may be present at shorter time scales. Indeed, $\rho_5(D^c,t)$ exhibits a succession of peaks in the first $0.7\seg$ of the discharge, when the C process evolves fastest. 
However, due to the temporal resolution of our sampling, it is hard to guarantee that these  are not just artifacts caused by fluctuations of the system.  

We will consider the origin of the fast-slow dynamics exhibited by the deformation processes in Sec.~\ref{sec::FT}. Before that, we will discuss the reversibility of noncollapse trajectories and develop necessary machinery for a quantitative comparison of different trajectories.

\subsection{Reversibility of noncollapse trajectories}
\label{sec::Reversibility}

At the end of a noncollapse process all visible deformation of the shell disappears. Thus the trajectories of noncollapse processes eventually return close to their initial condition.
This is corroborated by the fact that distances between the initial conditions of all trajectories are  smaller than the size of fluctuations. Since, hardware limitations restrict the duration of our sampling, we cannot observe the whole trajectory. However, we will demonstrate in this section how  to estimate the time when trajectories of noncollapse processes start returning. 
This will prove useful for comparing different trajectories in Secs.~\ref{sec::CrossCompare}~and~\ref{sec::FT}.

We start by considering the FFC trajectory $X^a$. 
Small values of $D^a$ in the rectangular region marked by a dotted line in Fig.~\ref{fig::matrices}(a) indicate that the states $X^a(t)$ for $t \in [6.00, 8.30]\seg$ are close to states visited previously, for $t \in [3.63, 4.23]\seg$.
Thus, the trajectory is retracing itself towards its initial condition.  This is not surprising as it reflects the reversible nature of the deformation. Nevertheless, it is remarkable that we can directly observe it by considering only a small portion of the deformation pattern.

The last two transitions of $X^a$ in Figs.~\ref{fig::matrices}(a)~and~(d) occur while the trajectory is returning to its initial condition. Accordingly, we call them \emph{backward} transitions.  All other observed transitions  occur as the deviation from the initial condition increases and we call them \emph{forward} transitions. 
Note that $X^a$ evolves slower as it returns. Also, note that the backward transitions are less abrupt than forward transitions and correspond to relatively small peaks of $\rho_{10}(D^a,t)$, see \protect Fig.~\ref{fig::matrices}(d).
Clearly, the trajectory $X^a$ starts returning to $X^a(0)$ at some time $t \in [4.27,5.97]\seg$, between the last forward and the first backward transitions.
This is consistent with the fact that $\Dev(X^a(t))$ reaches its maximum at $t = 4.84\seg$.
In general, fluctuations make it difficult to determine the precise time when a trajectory starts returning. However, the last forward and the first backward transitions can be identified with certainty for all FFC processes. The fact that the trajectories move very little between these transitions implies that the turning point of a trajectory $X$ is contained in a small ball around $X(t_r)$, where $t_r$ is the time of the last forward transition.  In the following, we refer to $t_r$ as the \emph{time of onset of return}. 

In the case of CTC processes our sample time is too short to capture the first backward transition.  For the trajectory $X^b$, Figs.~\ref{fig::matrices}(b)~and~\ref{fig::matrices}(e) show that after the transition at $t\approx 4\seg$ the system merely fluctuates during the rest of the sampled time. 
Given the total duration of the discharge, and the time the system spends fluctuating after the last observed transition, we expect that the trajectory does not evolve much further before it turns around. Thus, in analogy with the FFC case, we refer to the time of the last detected transition of a CTC trajectory as its time of onset of return, $t_r$. 

We expect the deformation to start receding after the effective axial stress on the shell starts decreasing.
In our experiments the granular discharge is triggered right after the filling of the silo.
Under these conditions, one expects that local frictional forces exerted by the grains will not satisfy the maximal mobilization condition~\cite{vanel99, perge12, cambau13, wang15}. Thus, the effective axial stress will increase at the beginning of the discharge as a result of a spontaneous mobilization process~\cite{perge12, cambau13}. 
Once the maximum possible stress component is transmitted to the shell, Janssen's saturating profile~\cite{janssen} is realized, and the axial stress starts decreasing. 
One can thus interpret $t_r$ as a rough estimate for the duration of the mobilization process. 

We remark that since $t_r$ is defined as the time of the last forward transition it represents only a lower bound on the time when the trajectory starts returning. A more precise determination of the return time would require the observation of the whole pattern. It is therefore not surprising that we find a large dispersion in the values of $t_r$. For FFC and CTC trajectories $t_r \in[2,4]\seg$ and there is no evidence of a dependence on their class or initial filling height. 
The values of $t_r$ are larger but of the same order of magnitude as the duration of the mobilization regime measured by~\cite{cambau13} for a system with comparable dimensions, except for  the shell thickness, and different boundary conditions. On the other hand, in~\cite{perge12} it has been observed that the duration of the mobilization process can be substantially longer than the values of $t_r$, as it is associated to the details of the granular flow pattern inside the silo.

\subsection{Comparing the trajectories}
\label{sec::CrossCompare}

So far we studied the dynamics of individual trajectories. In this section, we consider distances between all the sampled states to investigate relations between the trajectories for  different discharges. We show that the trajectories of all FFC processes lie in a narrow tube in the space of persistence diagrams. Therefore, the evolution of the deformation patterns for all FFC processes is topologically similar, although the  geometry of the patterns can vary significantly. Trajectories of CTC and C processes also start inside the narrow tube containing the FFC processes but exit it a short time after the onset of the discharge. Nevertheless, during this short time they traverse a large portion of the tube.

To construct the tube encompassing the FFC trajectories we use the point to set $d_{W^2}$ distance in $\Per$, defined as follows. Let $S \subset \Per$ be a set and $x$ a point in $\Per$, then the distance between $x$ and $S$ is given by 
\[
d(x,S) = \min \left\{d_{W^2}(x,y) \colon \,y\in S\right\}.
\]
For every FFC trajectory $X$ we define a tubular neighborhood with radius $q$ by 
\[
O_q(X) = \{x \in \Per \colon d(x,X([0,t_r])) < q \}. 
\]
Note that $O_q(X)$ is a neighborhood of the forward moving part of $X$, i.e., $X([0,t_r]) = \{X(t) \colon t \in [0,t_r]\}$.
Let $q_X$ be the smallest radius such that $O_{q_X}(X)$ contains all FFC trajectories. 
We choose the FFC trajectory $X^R$ for which  $q_{X^R} = \min\{ q_X \colon X \in \text{ FFC }\}$.
This gives  $q_{X^R} = 0.17$, which is of the same order as the estimated lower bound on the fluctuations in the system.
To simplify the notation we denote $O_{q_{X^R}}(X^R)$ by $O$.
We estimate the length of the tube $O$ using the arc length of the forward moving part of the trajectory $X^R$. 
We could define the arc length as the sum of distances between the consecutive sample points. However, to mitigate the artificial elongation of the trajectory caused by fluctuations, we consider displacements over a longer time scale $\tau = 0.43\seg$ (corresponding to $13$ time steps)  and  parametrize the trajectory by
\[
\ell(X(t_i)) = \frac{1}{13}\sum_{j=0}^i\Delta_{0.43}(X(t_j)).
\] 
Using this definition, the arc length of the forward moving part of the trajectory $X^R$ is $\ell(X^R(t_r)) = 1.37 > 8q_{X^R}$. 
We stress that all observed  FFC trajectories are contained in the relatively narrow tubular neighborhood  $O$ centered around $X^R$. Thus, all FFC trajectories closely follow $X^R$ and it is reasonable to use $X^R$ as a reference trajectory. 

\begin{figure}
\includegraphics{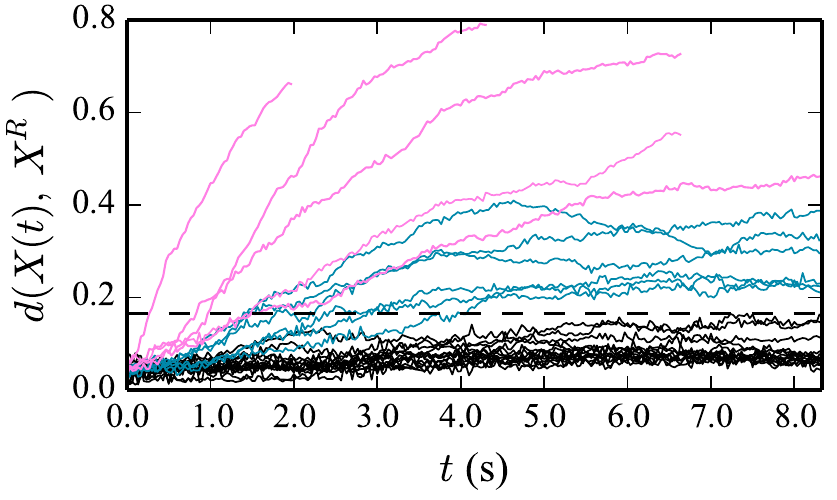}
\caption{Distance $d_{W^2}(X(t),X^R)$ between $X(t)$ and the reference trajectory $X^R$
 for all  FFC (black lines), CTC (blue lines) and C (pink lines) trajectories  in our sample.
The black dashed line marks the radius , $q_{X^R}$ of $O$. }
\label{fig::tube} 
\end{figure}

Figure~\ref{fig::tube} shows the deviation of all trajectories from the forward moving part of $X^R$.  Note that all the trajectories start close to $X^R$. By construction, every FFC trajectory  $X$ stays in $O$ and $d(X(t_i), X^R[0,t_r]) < 0.17$ for all $t_i$. 
However, C trajectories leave the set $O$ in less than $1.5\seg$ and they keep moving away from  $X^R$ until the collapse of the silo. CTC trajectories tend to stay in $O$ longer, some of them exit only after $4\seg$, and their distance to  $X^R$  eventually stops increasing as the pattern stabilizes.

We have shown that C and CTC trajectories start within the tube $O$ but they exit it very fast.
Finally, we identify what fraction of $O$ is traversed by C and CTC trajectories before they exit. To do this, we define the  arc length coordinate for any point in the tube  as follows.
Let  $x \in O$, then $\ell(x) = \ell(X^R(t_i(x)))$, where $t_i(x) = {\mbox{argmin}}_{t_j \in [0,t_r]}\{d_{w^2}(x,X^R(t_j))\}$. 
Suppose that   a trajectory $X$  exits the tube $O$ at \emph{exit time} $t_e$, i.e. 
$X(t) \in O$ for $0\leq t \leq t_e$ and  $X(t) \not\in O$ for $t > t_e$. Then $\ell(X(t_e))$ measures  the part of $O$ traversed by $X$ before exiting. 

For all C and CTC trajectories $\ell(X(t_e)) > 0.99$. This means that they follow closely a large part of $X^R$, so that initially they are similar to the FFC trajectories.
On the other hand, the fact that the C processes exit $O$ in such a short time indicates that, already at an early  stage of the discharge, their dynamics is very different from FFC processes. 
Note that $t_e$ for all C processes is smaller than the minimum value of $t_r$, our estimated lower bound for the duration of the mobilization process. 
This supports the idea, proposed in \cite{colonnello14, gutierrez15}, that the collapse of the silo is triggered by partially mobilized friction.
Moreover, the evolution of all C processes  is  fast compared to FFC processes. In a quarter of the time required by $X^R$ they traverse more than $70\%$ of its length. This suggests that comparing how fast the trajectories evolve at the onset of the discharge may allow for an early classification of the deformation processes. We explore this idea in Sec.~\ref{sec::classification}.

\subsection{Early collapse detection and the initial condition}
\label{sec::classification}

\begin{figure}
\includegraphics{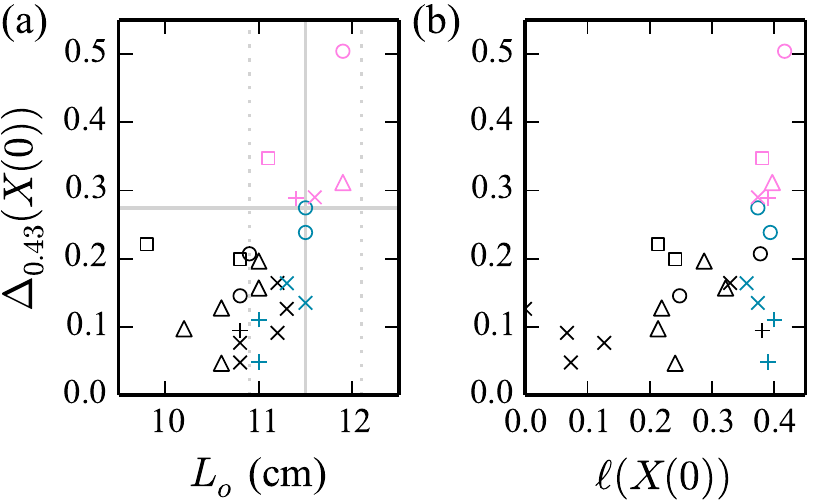}
\caption{For each trajectory $X$ in our sample, $\Delta_{0.43}(X(0))$ versus:
(a) Initial filling height of the corresponding granular discharge. The continuous (dotted) vertical gray line marks the critical filling height, $L_c$ (and its error bounds). The horizontal line marks a classification threshold for C trajectories. 
(b) $\ell(X(0))$.
All trajectories $X$ resulting from successive discharges of a single silo are represented with the same symbol, they are colored according to the class of $X$: FFC (black), CTC (blue) or C (pink).}
\label{fig::speed-TP} 
\end{figure}

Early prediction of silo collapse is an important problem in many industrial applications. As a toy example, 
we demonstrate a method that uses TDA for predicting the collapse in the case of our paper silos. 
The silo is known to collapse when its initial filling height exceeds the critical value $L_c$~\cite{nedderman, gutierrez09}.  
Thus one can expect that knowledge of $L_o$ and $L_c$ is enough to predict the outcome of the granular discharge. Unfortunately, as visible in Fig.~\ref{fig::speed-TP}(a), the error in $L_c$ is quite large and predicting the class of a deformation process based solely on $L_o$ leads to  misclassifications.

In Sec.~\ref{sec::CrossCompare} we showed that C processes evolve faster than noncollapse processes at the onset of the discharge. 
Thus, we propose a classifier based on the value of the initial displacement rate $\Delta_\tau(X(0))$. For  small values of $\tau$ the classifier does not yield satisfactory results because $\Delta_\tau(X(0))$ is dominated by noise. 
However, for time scales $\tau > 0.4\sec$, long enough to overcome noise, it performs consistently well. In particular, the value of $\Delta_{0.43}(X(0))$ allows us to distinguish collapse processes better than the value of $L_o$.

As shown in Fig.~\ref{fig::speed-TP}(a), for all C processes $\Delta_{0.43}(X(0)) > 0.28$, while for noncollapse processes $\Delta_{0.43}(X(0)) < 0.28$.  
It does not come as a surprise that for some CTC processes $\Delta_{0.43}(X(0))$ is very close to the threshold between these two classes, since according to our qualitative criteria CTC processes come very close to causing the collapse of the silo.  We stress that by using the value of $\Delta_{0.43}(X(0))$  we distinguish collapse and noncollapse processes already $0.43\seg$ after the onset of the discharge.  At this time all trajectories in our sample are inside the neighborhood $O$ and thus indistinguishable from an FFC trajectory through instantaneous observations. 

Figure~\ref{fig::speed-TP}(a) suggests that there exists a relationship between the initial rate of evolution, $\Delta_{0.43}(X(0))$, and $L_o$.  Although the dispersion of the results is fairly large, it is reduced when considering each silo separately. This suggests that some intrinsic properties of individual shells affect the dynamics under the discharge.  It is possible that microscopic defects of the paper sheet used to build a silo and  small asymmetries introduced during its construction are not detected by visual inspection. 
Such defects can make a shell more prone to develop initial deformations during filling. These are known to have an important effect on the resistance of the shell to axial compression~\cite{horton65, simitses86} and are thus likely to alter the dynamics of the deformation during the granular discharge. 
In particular, we expect $\Delta_{0.43}(X(0))$ to be largely determined by the initial deformations, which must be reflected in the initial condition $X(0)$.
 
To assess the effect of the initial condition, for each trajectory $X$ we consider the value of $\ell(X(0))$.  Figure~\ref{fig::speed-TP}(b) shows that $\Delta_{0.43}(X(0))$ tends to increase with $\ell(X(0))$.  So, a larger initial deformation leads to a faster evolution of the deformation pattern at the beginning of the discharge.  We expect that the large spread of the points in Fig.~\ref{fig::speed-TP}(b) is caused by the fact that we consider only a  small portion of the deformation pattern.  
We also note that  for the FFC processes $\ell(X(0))$ is smaller than for  C and CTC processes.  This suggests that the dynamics of the deformation pattern can be predicted from the initial condition. Thus, it is likely that by considering the initial deformation pattern on the whole surface of the silo one could distinguish collapse and noncollapse processes even before the discharge begins.

\subsection{Nature of the fast transitions}
\label{sec::FT}

\begin{figure}
\includegraphics{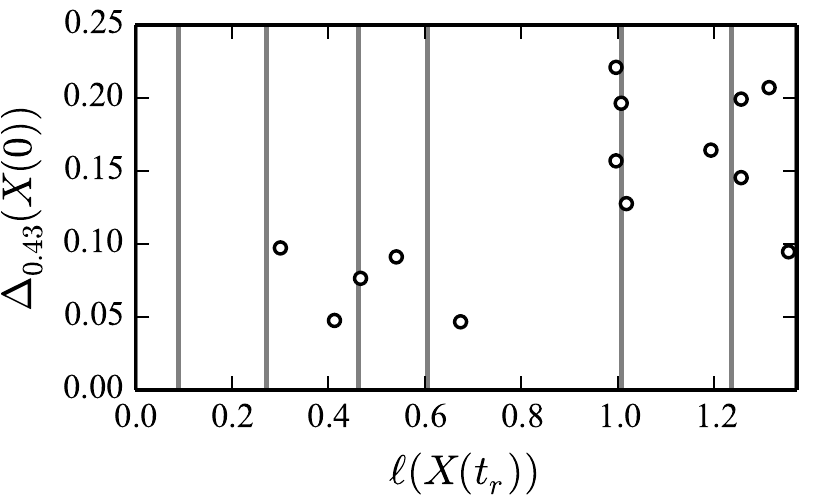}
\caption{Arc length coordinate of the points $X(t_r)$ for each FFC trajectory $X$ versus $\Delta_{0.43}(X(0))$. Vertical lines mark the coordinates of transition points of the trajectory $X^R$. }
\label{fig::t_r-on-R} 
\end{figure}

In this section we discuss the mechanism causing the fast-slow dynamics described in Sec.~\ref{sec::results::dynamics}. At first glance, the observed abrupt transitions might look surprising because the effective axial stress on the shell is expected to evolve rather smoothly~\cite{perge12, wang15}.
Naturally, fluctuations induced by the granular flow may affect the effective stress on the shell~\cite{bertho03}.
Moreover, both simulations~\cite{wang15, wang13} and direct measurements on large scale silos~\cite{zhong01, wang15} exhibit very large local stress fluctuations, particularly at the onset of the granular discharge. 
Thus, one possible explanation is that fast transitions in the evolution of the deformation are triggered by fluctuations induced by the flow.
However, in what follows we provide evidence that  stress fluctuations are not the main source of these abrupt transitions.   
We observe that transitions do not occur at random places along the tube $O$, as one would expect if they were predominately  caused by fluctuations. Thus, we argue that the slow-fast dynamics is intrinsic to the response of the shell during the granular discharge.

We concentrate only on the FFC trajectories because they are confined in the narrow tube $O$ around the trajectory $X^R$. Thus, the positions of their transitions can be meaningfully compared using the arc length coordinate of points in $O$ defined in Sec.~\ref{sec::CrossCompare}. We use the methods described in Sec.~\ref{sec::results::dynamics} to detect the transitions of FFC trajectories. As already mentioned, the last forward transitions can be identified with certainty for all FFC processes. Figure~\ref{fig::t_r-on-R}  shows their positions along the tube $O$. 
It is clear that they typically fall close to some transition of the trajectory $X^R$. 
This observation holds for $85\%$ of the transitions detected by our methods.
In the remaining cases, visual inspection of the distance matrices shows the detection of the transition time is not optimal, either due to fluctuations or because our sampling time is too long with respect to the time between two successive transitions.
Moreover, for  $50\%$ of the FFC trajectories all consecutive transitions fall close to consecutive transitions of $X^R$.  

The moderate amount of misaligned transitions suggests that the slow-fast dynamics is intrinsic to the behavior of the shell. This behavior is indeed reminiscent of the behavior of an empty cylindrical  shell subjected to axial compression. It has been experimentally observed~\cite{yamaki75} that, as the axial stress is increased, the empty shell suffers a sequence of successive transitions towards metastable states with increasing curvature. These metastable states are localized with respect to the length of the cylinder and periodic around its circumference. As suggested by~\cite{hunt2003cylindrical}, the circumferential wave number can be predicted  from the cylinder length. The lack of the characteristic axial symmetry in our experiments can be explained by a stabilizing effect of the flowing granular material~\cite{gutierrez15}.
In fact, one observes similar patterns when a close fitting rigid mandrel is fitted inside a thin cylindrical shell subject to axial compression~\cite{horton65, seffen14}, in this case the mandrel constrains the depth of any developing indentations.
However, the precise nature of any stabilization mechanism induced by the granular flow remains obscure and out of the scope of the present paper.

Now we return to Fig.~\ref{fig::t_r-on-R} and discuss the evident separation of the last forward transitions into two distinct groups. 
For trajectories with $\Delta_{0.43}(X(0)) < 0.10$ the value of $\ell(X(t_r))$ is generally smaller than $0.70$ and their last forward transition corresponds to one of the first four transitions of $X^R$. 
On the contrary, for all trajectories with $\Delta_{0.43}(X(0)) > 0.10$ the value of $\ell(X(t_r))$ is larger than $0.99$ and their last forward transition reaches at least the fifth transition of $X^R$.  
This suggests the presence of a bifurcation along the control parameter $\Delta_{0.43}(X(0))$, which describes how fast the deformation evolves at the onset of the discharge. Thus, we hypothesize that the transition exhibited by $X^R$ at $\ell(X^R(t_r)) = 1.01$ can only be triggered if the initial evolution of a deformation process is fast enough. Namely, if $\Delta_{0.43}(X(0))$ is larger than the critical value $0.10$.  Moreover, we propose that there is a second bifurcation along the parameter $\Delta_{0.43}(X(0))$ around the value $0.28$, that accounts for the separation between collapse and noncollapse trajectories observed in Fig.~\ref{fig::speed-TP}.    

We remark that we do not observe any bifurcation along $L_o$ and $\ell(X(0))$. We attribute this to the fact that they fail to capture some  important  aspect of the initial condition. As described in Sec.~\ref{sec::classification}, the effect of any initial defects of the shell cannot be accounted for by $L_o$. We also recall that we only observe a fraction of the silo, so $\ell(X(0))$ does not provide complete information about  the initial condition.  The parameter $\Delta_{0.43}(X(0))$  is also obtained only from a fraction of the shell, but the observed initial evolution of the deformation seems to be affected by the global state of the shell. Thus, it yields a more comprehensive description of the initial state of the system.

\section{Conclusion}
\label{sec::conclusions}

In this paper we have demonstrated that TDA constitutes an effective framework to investigate the complex deformation patterns sustained by thin walled silos during gravity driven granular discharges. Using TDA, a deformation process is associated with a trajectory in the space of persistence diagrams. This allows for a meaningful description of the time evolution of the deformation and a quantitative comparison of different processes. 

We found that a short time after the onset of the granular discharge, significant dynamical differences arise between those deformation processes that are reversible and those that lead to the collapse of the silo. 
By this time, the trajectories representing all the deformation processes are very close, but the deformation develops much faster in processes that lead to the collapse of the silo. 
Based on these observations, we  proposed an early classification criterion that distinguishes processes that lead to a collapse while the deformation of the shell is still incipient. 
This is remarkable considering the low temporal and spatial resolution of our sample, and the fact that we observe only a small portion of the pattern.
The efficacy of this toy-model prediction method for the collapse demonstrates the strong potential of our methods for designing early-warning signals.  
Moreover, recent progress in computational homology makes it possible to detect the differences in the dynamics of the processes in real time, even for considerably  larger data sets than considered in this paper.

We also showed that reversible deformation processes are characterized by slow-fast dynamics. 
Our results suggest that this is associated to the intrinsic nonlinear behavior of the shell, evocative of the sequence of post buckling deformation patterns observed on empty cylindrical shells subject to an increasing axial compression.
However, additional experimental observations are needed to study this phenomenon. 
In particular, we found that in some cases our sampling rate was too slow to detect accurately the transitions between intervals of slow evolution.
Specially for deformation processes that lead to the collapse of the silo, for which our sampling rate does not allow to conclusively decide if slow-fast dynamics is present.

\section{Acknowledgements}
The authors would like to thank Konstantin Mischaikow for many fruitful discussions and the use of his cluster computer Conley3. 
They also express their gratitude to Lou Kondic for very useful comments on the manuscript.
CC would like to acknowledge the late Gustavo Guti\'errez, who contributed with many valuable ideas and the laboratory where experiments were performed.
Efficient algorithms implemented in the GUDHI library were used to compute the persistence diagrams.
MK was supported by ERC project GUDHI (Geometric Understanding in Higher Dimensions).

\bibliographystyle{unsrt}

\begin{thebibliography}{10}

\bibitem{rotter08}
J.~M.~Rotter,
\newblock {\em Bulk Solids Handling: Equipment Selection and Operation}, 
  (2008) p. 99.

\bibitem{dogangun09}
A.~Dogangun, Z.~Karaca, A.~Durmus, and H.~Sezen,
\newblock {\em J. Perform. Constr. Fac.}, {\bf 23}, 65 (2009).
  
\bibitem{dutta13}
A.~B.~Dutta, 
\newblock {\em Global Research Analysis}, {\bf 2}, 41 (2013).


\bibitem{liu95}
C.~H.~Liu, S.~R.~Nagel, D.~A.~Schecter, S.~N.~Coppersmith, T.~S.~Majumdar, O.~Narayan, and T.~A.~Witten,
\newblock {\em Science}, {\bf 269}, 513 (1995).

\bibitem{majmudar05}
T.~S.~Majmudar, and R.~P.~Behringer,
\newblock {\em Nature}, {\bf 435} 1079 (2005).

\bibitem{vanel99b}
L.~Vanel, D.~Howell, D.~Clark, R.~P.~Behringer, and E.~Cl{\'e}ment,
\newblock {\em Phys. Rev. E}, {\bf 60} (1999).

\bibitem{janssen}
H.~A.~Janssen,
\newblock {\em Zeitschr. d. Vereines deutscher Ingenieure}, {\bf 39} 1045 (1895).

\bibitem{nedderman}
R.~M.~Nedderman,
\newblock {\em {Statics and kinematics of granular materials}}.
\newblock (Cambridge University Press, 1992).

\bibitem{vanel99}
L.~Vanel, and E.~Cl{\'e}ment,
\newblock {\em Eur. Phys. J. B}, {\bf 11} 525 (1999).

\bibitem{ovarlez03}
G.~Ovarlez, C.~Fond, and E.~Cl{\'e}ment,
\newblock {\em Phys. Rev. E}, {\bf 67}, 060302 (2003).

\bibitem{bertho03}
Y.~Bertho, F.~Giorgiutti-Dauphin{\'e}, and J.~P.~Hulin,
\newblock {\em Phys. Rev. Lett.}, {\bf 90}, 144301 (2003).

\bibitem{perge12}
C.~Perge, M.~A.~Aguirre, P.~A.~Gago, L.~A.~Pugnaloni, D.~{Le Tourneau}, and
  J.~C.~Geminard,
\newblock {\em Phys. Rev. E}, {\bf 85}, 021303 (2012).

\bibitem{cambau13}
T.~Cambau, J.~Hure, and J.~Marthelot,
\newblock {\em Phys. Rev. E}, {\bf 88}, 022204 (2013).

\bibitem{wang15}
Y.~Wang, Y.~Lu, and J.~Y.~Ooi,
\newblock {\em Powder Technol.}, {\bf 282}, 43 (2015).

\bibitem{back11}
R.~Back,
\newblock {\em Granul. Matter}, {\bf 13}, 723 (2011).

\bibitem{wang13}
Y.~Wang, Y.~Lu, and J.~Y.~Ooi,
\newblock {\em Eng. Struct.}, {\bf 56}, 1308 (2013).

\bibitem{zhong01}
Z.~Zhong, J.~Y.~Ooi, and J.~M.~Rotter,
\newblock {\em Eng. Struct.}, {\bf 23}, 756 (2001).

\bibitem{ostendorf03}
M.~Ostendorf, J.~Schwedes, J.~Bohrnsen, and H.~Antes,
\newblock {\em Task Quarterly}, {\bf 7}, 611 (2003).

\bibitem{horton65}
W.~H.~Horton, and S.~C.~Durham,
\newblock {\em Int. J. Solids. Struct.}, {\bf 1}, 59 (1965).

\bibitem{simitses86}
G.~J.~Simitses,
\newblock {\em Appl. Mech. Rev.}, {\bf 39}, 1517 (1986).

\bibitem{jansseune16}
A.~Jansseune, W.~De~Corte, and J.~Belis,
\newblock {\em Int. J. Solids. Struct.}, {\bf 96}, 92 (2016).

\bibitem{skukis17}
E.~Skukis, O.~Ozolins, K.~Kalnins, and M.~A.~Arbelo,
\newblock {\em Procedia Eng.}, {\bf 172}, 1023 (2017).

\bibitem{virot17}
E.~Virot, T.~Kreilos, T.~M.~Schneider, and S.~M.~Rubinstein,
\newblock {\em Phys. Rev. Lett.}, {\bf 119}, 224101 (2017).

\bibitem{gutierrez15}
G.~Guti{\'e}rrez, C.~Colonnello, P.~Boltenhagen, J~.R.~Darias, R.~Peralta-Fabi, F.~Brau, and E.~Cl{\'e}ment,
\newblock {\em Phys. Rev. Lett.}, {\bf 114}, 018001 (2015).

\bibitem{colonnello14}
C.~Colonnello, L.~I.~Reyes, E.~Cl\'ement, and G.~Guti\'errez,
\newblock {\em Physica A}, {\bf 398}, 35 (2014).

\bibitem{gallego15}
E.~Gallego, A.~Ruiz, and P.~J.~Aguado,
\newblock {\em Comput. Electron. Agric.}, {\bf 118}, 281 (2015).

\bibitem{zhao13}
C.~Zhao, H.~Matsuda, S.~Lou, C.~Morita, and A.~Koga,
\newblock {\em Appl. Math. Inf. Sci.}, {\bf 7}, 999 (2013).

\bibitem{edelsbrunner-harer}
H.~Edelsbrunner, and J.~L.~Harer,
\newblock {\em Computational topology: an introduction}
\newblock (American Mathematical Society, 2010).

\bibitem{carlsson}
G.~Carlsson.
\newblock {\em Bull. Amer. Math. Soc. (N.S.)}, {\bf 46}, 255 (2009).

\bibitem{PhysD2016}
M.~Kr\'amar, R.~Levanger, J.~Tithof, B.~Suri, M.~Xu, M.~Paul, M.~Schatz, and K.~Mischaikow.
\newblock {\em Physica D}, {\bf 334}, 82 (2016).

\bibitem{yoshimura55}
Yoshimura Yoshimaru,
\newblock NACA Technical Memorandum 1390, 1955.


\bibitem{pini16}
V.~Pini, J.~J.~Ruz, P.~M.~Kosaka, O.~Malvar, M.~Calleja, and J.~Tamayo,
\newblock {\em Sci. Rep.}, {\bf 6} (2016).

\bibitem{kondic2017evolution}
L.~Kondic, M.~Kram{\'a}r, L.~Koval{\v{c}}inov{\'a}, and K.~Mischaikow,
\newblock In {\em EPJ Web of Conferences}, {\bf 140}, 15014 (2017).

\bibitem{fisher1936use}
R.~A.~Fisher,
\newblock {\em Ann. Hum. Genet.}, {\bf 7}, 179 (1936).


\bibitem{gutierrez09}
G.~Guti{\'e}rrez, P.~Boltenhagen, J.~Lanuza, and E.~Cl{\'e}ment,
\newblock In {\em {Traffic and Granular Flow'07}} (2009) p. 517.


\bibitem{yamaki75}
N.~Yamaki, K.~Otomo, and K.~Matsuda,
\newblock {\em Exp. Mech.}, {\bf 15}, 23 (1975).

\bibitem{hunt2003cylindrical}
G.~W.~Hunt, G.~J.~Lord, and M.~A.~Peletier,
\newblock {\em Discrete Cont. Dyn. B}, {\bf 3}, 505 (2003).

\bibitem{seffen14}
K.~A.~Seffen, and S.~V.~Stott,
\newblock {\em J. Appl. Mech. T. ASME}, {\bf 81}, 061001 (2014).


\end{thebibliography}

\end{document}